\documentclass[aps,prl,twocolumn,superscriptaddress,footinbib]{revtex4-1}  
\usepackage{graphicx}  
\usepackage{dcolumn}   
\usepackage{bm}        
\usepackage{amssymb}   
\usepackage{amsmath}   
\usepackage{extarrows}

\newcommand{\bra}[1]{{\left\langle  #1 \right|}}
\newcommand{\ket}[1]{{\left|  #1 \right\rangle}}

\newcommand{\hide}[1]{}

\newcommand{\bn}{{\bm n}}

\newcommand{\be}{\begin{equation}}
\newcommand{\ee}{\end{equation}}
\newcommand{\bes}{\begin{eqnarray}}
\newcommand{\ees}{\end{eqnarray}}
\begin{document}

\title{Floquet-heating-induced Bose condensation in a scar-like mode of an open driven optical-lattice system}
\author{Alexander~Schnell} 
\email[Electronic address: ]{schnell@tu-berlin.de}
\affiliation{Technische Universit{\"a}t Berlin, Institut f{\"u}r Theoretische Physik, 10623 Berlin, Germany}
\author{Ling-Na~Wu} 
\affiliation{Technische Universit{\"a}t Berlin, Institut f{\"u}r Theoretische Physik, 10623 Berlin, Germany}
\author{Artur Widera}
\affiliation{Department of Physics and State Research Center OPTIMAS, Technische Universit\"at Kaiserslautern, Germany}\author{Andr{\'e}~Eckardt} 
\affiliation{Technische Universit{\"a}t Berlin, Institut f{\"u}r Theoretische Physik, 10623 Berlin, Germany}

\date{\today}

\begin{abstract}
Periodically driven quantum systems suffer from heating via resonant excitation. While such Floquet heating guides a generic isolated 
system towards the infinite-temperature state, a driven \emph{open} system, coupled to a thermal bath, will approach a non-equilibrium 
steady state. We show that the interplay of bath-induced dissipation and controlled Floquet heating can give rise to non-equilibrium Bose condensation in a mode protected from Floquet heating. In particular, we consider a one-dimensional (1D) Bose gas in an optical lattice of finite extent, which is coupled weakly to a three-dimensional thermal bath given by a second atomic species. The bath temperature $T$ lies well above the crossover temperature, below which the majority of the system's particles form a (finite-size) Bose condensate in the ground state. However, when a strong local potential modulation is switched on, which resonantly excites the system, a non-equilibrium Bose condensate is formed in a state that decouples from the drive. Our predictions, which are based on a microscopic model that is solved using kinetic equations of motion derived from Floquet-Born-Markov theory, can be probed under realistic experimental conditions. 
\end{abstract}

\maketitle

\emph{Introduction.}---
Floquet engineering is a powerful tool for controlling isolated quantum systems by means of time-periodic forcing \cite{GoldmanDalibard14, BukovEtAl15, Eckardt17, OkaSota19}. Prominent examples include the control of phase transitions \cite{EckardtEtAl05b, ZenesiniEtAl09, StruckEtAl13, StraeterEckardt15, SongEtAl22}, the engineering of artificial magnetic fields and topological band structures in systems of charge-neutral particles, such as atoms or photons \cite{OkaAoki09, AidelsburgerEtAl11, StruckEtAl12, AidelsburgerEtAl13, RechtsmanEtAl13, JotzuEtAl14, TaiEtAl17, WangEtAl18}, as well as the realization of so-called anomalous Floquet topological states that cannot exist in undriven systems \cite{KitagawaEtAl10, RudnerEtAl13,MaczzewskyEtAl17, WinterspergerEtAl20}. Nevertheless, Floquet (i.e.\ periodically driven quantum) systems also suffer from heating, as it is caused by unwanted resonant excitation processes \cite{WeinbergEtAl15, AbaninEtAl15, BilitewskiCooper15b, Straeter_16, ReitterEtAl17, SunEckardt2020, ViebahnEtAl21}. 
Such Floquet heating will generically guide an isolated system towards an infinite-temperature state, corresponding to eigenstate thermalization without energy conservation \cite{DAlessioRigol14,LazaridesEtAl14b}. 
However, when a Floquet system is coupled to a bath \cite{BluemelEtAl91,KohlerEtAl97,BreuerEtAl00, HoneEtAl09, KetzmerickWustmann10, VorbergEtAl13, LangemeyerHolthaus14, szczygielski2014, VorbergEtAl15, HaddadfarshiEtAl15, DaiEtAl16, ShiraiEtAl16, RestrepoEtAl17, SchnellEtAl20, IkedaEtAl21, SchnellEtAl21}, it will not approach infinite temperature, but a non-equilibrium steady state.

Previous work addressed the preparation of equilibrium-like states in open Floquet systems \cite{ShiraiEtAl14,DehghaniEtAl15, SeetharamEtAl15, IadecolaEtAl15, QinEtAl18}. 
Here, we show that Floquet heating can be exploited for robust preparation of interesting non-equilbrium states. Namely, 
we consider a Bose gas in contact with a thermal bath of temperature well above the critical temperature and find that the heating induced by strong resonant driving can make the system Bose condense into an excited state, which is decoupled from the drive. In the following, we discuss this intriguing effect using a realistic microscopic model describing a two-species mixture of ultracold atoms \cite{SpethmannEtAl12, HohmannEtAl17, SchmidtEtAl19PSS, BoutonEtAl20,SchmidtEtAl18, SchmidtEtAl19,Burchianti2020,MilEtAl20, ZhuEtAl19,LauschEtAl18, LenaDaley2019, KleinEtAl07, ostmann2017cooling, LauschEtAl18, LauschEtAl18_2, RammohanEtAl21}. We solve this model numerically using kinetic equations and stochastic equations \cite{VorbergEtAl13, VorbergEtAl15} based on Floquet-Born-Markov theory \cite{BluemelEtAl91, KohlerEtAl97, HoneEtAl09} and provide an intuitive explanation of the effect.

\begin{figure}
\centering
\includegraphics[scale=0.95]{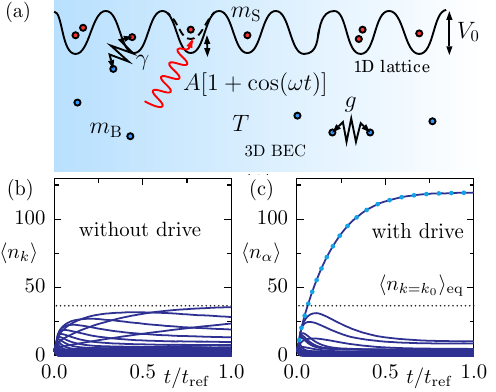}
\caption[]{
(a)~ System: $N=200$ non-interacting bosons in a 1D optical lattice of $M=49$ sites 
subject to local time-periodic potential modulations of amplitude $A$, frequency $\hbar \omega=1.5J$ at site $\ell=20$. Bath: Weakly interacting 3D Bose condensate of temperature $T\approx2.38J$ well above the crossover temperature $T_c^\text{eq}\approx 0.7 J$ below which the majority of lattice bosons occupies the ground state in absence of driving. 
(b,c) Monte-Carlo simulation of the relaxation dynamics of the mean occupations of (b) the single-particle eigenstates for the undriven system with $A=0$ and (c) the single-particle Floquet states of the driven system with $A=J$. Statistical errors are smaller than the linewidth and therefore not shown. Dotted line: thermal ground-state occupation, blue bullets: occupation $ \langle n_{k_c}\rangle$ of the undriven eigenstate $k_c=2\pi/(\ell a)$.}
\label{fig:main}
\end{figure}

\emph{The System.}---%
We consider $N$ noninteracting bosonic atoms in a 1D optical lattice with strong transverse confinement, 
described by the tight-binding Hamiltonian
\begin{equation}
\hat{H}_0= -J\sum_{i=1}^{M-1} ( \hat{a}_{i+1}^\dag \hat{a}_i + \hat{a}_{i}^\dag a_{i+1}) = \sum_k \varepsilon_k \hat{b}^\dag_k \hat{b}_k,
\label{eq:tight-bind}
\end{equation}
with tunnelling parameter $J$, number of lattice sites $M$ (assuming box-type confinement) and annihilation operator $\hat{a}_i$ for a boson on site~$i$. The eigenmodes, with annihilation operators $\hat b_k =\sum_i \langle \psi_k | i \rangle \hat a_i$, are characterized by wavefunctions $\langle i|\psi_k\rangle = \sqrt{2/(M+1)} \sin(k ai)$ and energies $\varepsilon_k=-2J\cos(ka)$, with lattice spacing $a$ and wavenumbers $k=\nu \pi/[a(M+1)]$, with $\nu=1,\ldots, M$.
Additionally, the system is subjected to a local time-periodic potential modulation of amplitude $A$ and frequency $\omega$ on site $\ell$ [cf.~Fig.~\ref{fig:main}(a)], giving the total Hamiltonian
\begin{equation}
\hat{H}(t)= \hat H_0 + \hat H_D(t), \quad \hat{H}_D(t)= A[1+\cos(\omega t)]\hat a^\dag_\ell \hat a_\ell . 
\end{equation}
The box confinement and the local modulation can be realized using spatial light modulators. 
The single-particle Floquet states, $|\varphi_\alpha(t)\rangle = e^{-i \varepsilon_\alpha t/\hbar}|u_\alpha(t)\rangle$, labeled by $\alpha=1,\ldots, M$, are characterized both by the quasienergies $\varepsilon_\alpha$ and the time-periodic Floquet modes
$|u_\alpha(t)\rangle=|u_\alpha(t+\mathcal{T})\rangle$, with driving period $\mathcal{T}=2\pi/\omega$. They are 
the eigenstates with eigenvalues $\exp(-i\varepsilon_\alpha \mathcal{T}/\hbar)$ of the single-particle one-cycle evolution operator from time $t$ to $t+\mathcal{T}$.
We also define the Floquet-mode annihilation operators 
$\hat{f}_\alpha=\sum_i \langle u_\alpha(t)| i \rangle \hat{a}_i$.

\emph{The Bath.}---%
The system interacts weakly with a bath given by a weakly interacting 3D Bose-Einstein condensate (BEC) of another atomic species. The bath temperature lies well below the bath's critical temperature, 
but well above the crossover temperature $T_c^\text{eq}$ below which the majority of lattice bosons occupies the ground state in absence of driving [cf.~Fig.~\ref{fig:main}(a)]. 
Similar scenarios have recently been realized experimentally   \cite{SchmidtEtAl18, SpethmannEtAl12,HohmannEtAl17,SchmidtEtAl19PSS,BoutonEtAl20,SchmidtEtAl19,MilEtAl20,ZhuEtAl19,Burchianti2020}.  
Approximating the bath as homogeneous with number density $n_\mathrm{B}$ over the extent of the lattice system
 and applying standard Bogoliubov theory \cite{PethickSmith}, the bath Hamiltonian reads
$
\hat{H}_\mathrm{B}=\sum_{\vec{q}}  E_\mathrm{B}(q) \hat{\beta}^\dagger_{\vec{q}}\hat{\beta}_{\vec{q}}
$.
Here   $E_\mathrm{B}(q) =  \sqrt{E_0^2(q) + 2  G E_0(q)}$ is the energy and 
$\hat{\beta}_{\vec{q}}$ the annihilation operator of a Bogoliubov quasiparticle with momentum $\hbar \vec{q}$, with $q=|\vec{q}|$, 
 $E_0(q)=\hbar^2q^2/{(2m_\text{B})}$, bare mass $m_\text{B}$, $G=g n_\text{B}$, and intrabath contact interaction strength $g$. 

The system particles interact with the bath particles via contact interactions of strength $\gamma$, described by
$
\hat{H}_\mathrm{SB}
= \gamma \int \!\mathrm{d}^3\vec{r}\, \hat{n}_\text{S}(\vec{r})\hat{B}(\vec{r})
$,
and
$
\hat{B}(\vec{r})=[\hat{n}_\text{B}(\vec{r})- n_\text{B}]
$,
where $\hat{n}_\text{S}(\vec{r})=\sum_{ij} w_i(\vec{r})w_j^*(\vec{r})\hat a^\dag_i \hat a_j$, with Wannier function $w_i(\vec{r})=\langle \vec{r}|i\rangle$ of lattice site $i$, describes the density of the system particles and $\hat{n}_\text{B}(\vec{r})$ the density of (bare) bath particles.
Subtracting the mean density $n_\mathrm{B}$, corresponding to an irrelevant energy shift, ensures that $\mathrm{Tr}_\mathrm{B}(\hat \varrho_\mathrm{B} {\hat{H}}_\mathrm{SB})=0$, as required by the Born-Markov formalism \cite{SM}. The bath state is given by $\hat \varrho_\mathrm{B}=\exp(-\hat H_\mathrm{B}/T)/Z_\mathrm{B}$ (with $k_\mathrm{B}\equiv1$). In leading (linear) order with respect to the Bogoliubov modes (describing single-phonon scattering, which is dominant for low bath temperatures \cite{LauschEtAl18}), 
$\hat{B}(\vec{r})\simeq\sqrt{n_B/V}\sum_{\vec{q}\ne \vec{0}}e^{i\vec{q}\cdot\vec{r}}[ u_q\hat{\beta}_{\vec{q}}-v_q\hat{\beta}_{-\vec{q}}^\dag]$, where $u_q$ and $v_q$, with $u_q^2- v_q^2=1$ and $2 u_q v_q=G/E_B(q)$,  are the usual Bogoliubov coefficients \cite{SM} and volume $V$ of the bath.

We assume ultraweak system-bath coupling, which is small compared to all single-particle (quasi)energy splittings in the system (so that also bath-mediated interactions are negligible). Under this assumption, we derive a master equation using Floquet-Born-Markov 
theory in secular approximation
\cite{BreuerPetruccione,KohlerEtAl97,BreuerEtAl00,HoneEtAl09,Wustmann10,DiermannEtAl19}. 
Then, the off-diagonal matrix elements of the density matrix $\rho$ in Floquet-state representation decouple from the diagonal ones and decay rapidly \cite{BreuerEtAl00,HoneEtAl09,Wustmann10}, so that $\hat \varrho(t)\simeq\sum_\mathbf{n} p_\mathbf{n}(t) \ket{\bn}\bra{\bn}$. Here $p_\bn(t)$ is the probability of the system for being in the many-body Floquet state $\ket{\bn}\equiv\ket{\bn(t)}=\ket{\bn(t+\mathcal{T})}$, characterized by the vector of occupation numbers $\bn=(n_1, \ldots, n_M)$ for single-particle Floquet modes $|u_\alpha(t)\rangle$. 
The probabilities follow the classical many-body rate equations \cite{VorbergEtAl15}
\be
\dot{p}_\bn(t) 
= \sum_{\alpha \beta} (1+n_\beta) n_\alpha \big[R_{\alpha \beta}p_{\bn_{\beta\leftarrow \alpha}}(t)-R_{\beta\alpha}p_\bn(t)\big],
\label{eq:rateq-fock}
\ee
where $\bn_{\beta\leftarrow \alpha}$ are occupation numbers obtained from $\bn$ by 
transferring a particle from mode $\alpha$ to $\beta$ and $R_{\alpha \beta}$ denotes the corresponding golden-rule-type single-particle rate
\begin{align}
	R_{\alpha \beta} = \frac{2\pi \gamma^2}{\hbar} \sum_{K\in \mathbb{Z}} \sum_{ij} (v_i)^{(K)*}_{\alpha \beta}(v_j)^{(K)}_{\alpha \beta} W_{ij}(\Delta^{(K)}_{\alpha \beta}).
	\label{eq:rates-general}
\end{align}
Here $(v_i)^{(K)}_{\alpha \beta}= \mathcal{T}^{-1} \int_0^{\mathcal{T}}\! \mathrm{d}t \,\mathrm{e}^{-iK\omega t} \langle u_\alpha(t)|i\rangle\langle i|u_\beta(t)\rangle$ are matrix elements and $W_{ij}(E)=J_{ij}(E)/({\mathrm{e}^{E/ T}-1})$ bath correlation functions, with spectral densities $ {J}_{ij}(E) =\mathrm{sgn}(E)  n_\mathrm{B}  {q(E)^3}I_{ij}(q(E))/(8{\pi}^2{\sqrt{E^2 + G^2}})$, wavenumber $q(E)$ for a bath quasiparticle with energy $E$, and $I_{ij}(q) \approx \mathrm{e}^{-\frac{1}{2} q^2 d^2} 2 \mathrm{sinc}[qa(i-j)]$ (obtained by approximating Wannier functions by oscillator ground states with isotropic oscillator length $d$)  \cite{SM}. 

\begin{figure}
	\centering
	\includegraphics[scale=0.95]{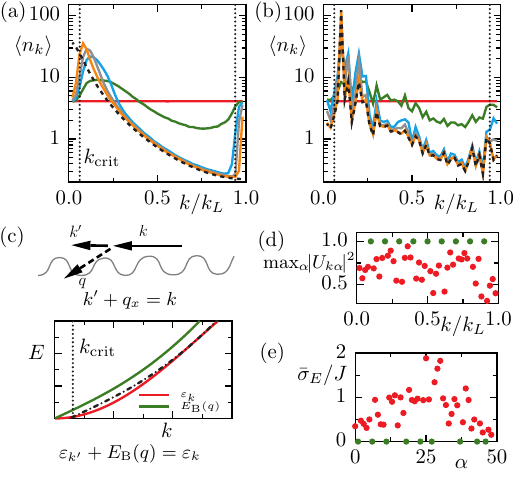}
	\caption[]{(a, b)~Snapshots of the momentum distribution $\langle n_k\rangle$  corresponding to Fig.~\ref{fig:main}(b, c), respectively, at times $t/t_\mathrm{ref}=0, 0.01, 0.05, 0.2, 1$ (red, green, blue, grey, orange). Dashed line: steady-state distribution inferred from the mean-field theory, dotted lines: critical momenta $k_\mathrm{crit}$ below (above) which one-phonon scattering is suppressed in the undriven case. (c) Top: Collision leading to the emission of a bath phonon. Bottom:  Bogoliubov dispersion of the bath (green) and lattice dispersion (red). Only after shifting the bath dispersion (dash dotted) by $k'\geq k_\mathrm{crit}$ there are multiple intersections. (d) Largest overlap $U_{k\alpha} = \langle \psi_k| u_\alpha(0)\rangle$ between the eigenstates $k$ of the undriven and the Floquet states $\alpha$ of the driven system. 
	(e) Cycle-averaged standard deviation $\bar \sigma_E^2 =\mathcal{T}^{-1}\int_0^\mathcal{T} \bra{u_\alpha(t)} \hat{H}^2(t) \ket{u_{a}(t)}\mathrm{d}t - E_{\mathrm{avg}, \alpha}^2$ of the energy $E_{\mathrm{avg}, \alpha}=\mathcal{T}^{-1}\int_0^\mathcal{T} \bra{u_\alpha(t)} \hat{H}(t) \ket{u_{a}(t)}\mathrm{d}t$ of the Floquet mode $\alpha$. (d,e) Undriven modes colored green.}
	\label{fig:dyn-occ}
\end{figure}

Equations (\ref{eq:rateq-fock}) describe exponentially many probabilities $p_\bn(t)$, but single- and few-particle expectation values, like
 $\langle \hat{n}_\alpha\rangle(t)$  or $\langle \hat{n}_\alpha\hat{n}_\beta\rangle(t)$ with $\hat{n}_\alpha=\hat{f}_\alpha^\dagger \hat{f}_\alpha$, can be obtained efficiently by quantum-jump Monte-Carlo simulations \cite{PlenioKnight98}, i.e.\ by sampling over random walks (trajectories) between different Fock states $\ket{\bn}$ \cite{VorbergEtAl15}. This method gives 
quasiexact results, in the sense that the accuracy is controlled by the number of trajectories. 
Alternatively, we can obtain  $\langle \hat{n}_\alpha\rangle(t)$ from the equations of motion
\cite{VorbergEtAl13,VorbergEtAl15,SchnellEtAl18}
$
	\mathrm{d} {\langle {\hat{n}}_\alpha \rangle}/\mathrm{d}t = \sum_\beta \left[ R_{\alpha \beta}  \langle (\hat{n}_\alpha+1) \hat{n}_\beta \rangle - R_{\beta\alpha}  \langle (\hat{n}_\beta+1) \hat{n}_\alpha \rangle  \right].
	\label{eq:Rateq-ex}
$
These depend, however, on two-body correlations $ \langle \hat{n}_\alpha \hat{n}_\beta \rangle$, which reflects that, even though we assume vanishing intra-system interactions, we are still dealing with an interacting problem, since the coupling operator 
$\hat{H}_\mathrm{SB}$ is cubic in the system and bath (quasiparticle) operators. We obtain a closed set of non-linear kinetic equations for the mean occupations $\langle \hat{n}_\alpha\rangle(t)$ by additionally making the mean-field approximation
 $\langle \hat{n}_\alpha\hat{n}_\beta\rangle \approx \langle \hat{n}_\alpha\rangle\langle\hat{n}_\beta\rangle$ (corresponding to a Gaussian ansatz for the system state) \cite{SM,VorbergEtAl15}. In the following, we mainly show exact Monte-Carlo solutions of Eq.~(\ref{eq:rateq-fock}). However, when scannig large parameter spaces for steady states, we employ also the kinetic equations, which still provide an excellent approximation. This can be seen from Fig.~\ref{fig:dyn-occ}(b), where the orange solid line, showing the late-time momentum distribution of the driven system, agrees very well with the black dashed line, giving the steady state prediction of the kinetic theory. 
 

 \emph{Parameters.}---
Inspired by recent experiments  \cite{SchmidtEtAl18, SpethmannEtAl12,HohmannEtAl17,SchmidtEtAl19PSS,BoutonEtAl20,SchmidtEtAl19,MilEtAl20,Burchianti2020}, we assume $N=200$ bosonic $^{39}$K atoms on $M=49$ lattice sites immersed in a bath of $^{87}$Rb (and discuss results for Cs in Rb in the supplemental material \cite{SM}). We consider a lattice depth of  $V_0=6 E_\text{R}$, with recoil energy $E_\text{R}=\hbar^2k_L^2/(2m_\mathrm{S})$, Potassium mass $m_\mathrm{S}$, lattice momentum $k_L=\pi/a$ and lattice spacing $a=395.01\mathrm{nm}$ corresponding to the Rb tune-out wavelength of $790.01\mathrm{nm}$ \cite{LeBlancThywissen2007,SchmidtEtAl2016}. 
For convenience, the lattice minima are considered to be isotropic, which slightly underestimates the transverse confinement. Moreover, the bath particles interaction parameter reads $G=g n_\mathrm{B}=0.05 E_\mathrm{R}$, (corresponding to $g = 2\pi\hbar^2a_\mathrm{Rb} /m_\mathrm{B}$, $a_\mathrm{Rb} = 104$ Bohr radii, and $n_\mathrm{B}=6.29 / a^3$).
Within our theoretical framework, the system--bath coupling $\gamma$ enters through the time scale 
$t_\mathrm{ref}=16{\pi \hbar^3}/({m_\mathrm{B}k_L n_\mathrm{B}\gamma^2})$ of the relaxation dynamics. In an experiment  $\gamma = 2\pi\hbar^2a_\mathrm{SB} /\tilde m$ is given by the K-Rb scattering length $a_\mathrm{SB}$ and reduced mass $\tilde m$. 
The bath possesses the temperature $T=2.38J$, corresponding to 15 percent of its estimated critical temperature $T_c^\text{Bath}=15.9J$ and the system is initialized in an infinite-temperature state (within the lowest band), with all Floquet/eigen modes populated equally. 
 
  \emph{Relaxation dynamics of the undriven system.}---
 Let us first discuss the equilibration dynamics of the undriven system, $A=0$. 
 Here the Floquet modes and quasienergies equal the single-particle eigenstates of $\hat{H}_0$ and their energies. In Fig.~\ref{fig:main}(b) we show the time-evolution of their mean occupations $\langle \hat{n}_k \rangle=\langle \hat{b}^\dag_k\hat{b}_k \rangle$ using the Monte-Carlo simulations. Since the system is one-dimensional, equilibrium Bose condensation is a finite-size effect. The corresponding crossover temperature, below which the majority of bosons occupies the single-particle ground state and the coherence length exceeds the system extent, can be estimated as $T_c^\mathrm{eq} \approx  8.3 J N/M^2\approx 0.69 J$ \cite{SchnellEtAl17, SM}. As the bath temperature $T=2.38J$ lies above this crossover temperature, at equilibrium, Bose condensation is not expected. This is confirmed in Fig.~\ref{fig:main}(b), where the dotted line indicates the thermal occupation of the ground state, which is approached in the long-time limit.


We, moreover, observe that a rather fast dynamics for $t \lesssim 0.1 t_\mathrm{ref}$ is followed by very slow relaxation taking much longer than $t_\text{ref}$. The reason for this separation of time scales becomes apparent from Fig.~\ref{fig:dyn-occ}(a) where we depict snapshots of the distribution $\langle\hat n_k \rangle$ at intermediate times~$t$ (solid lines) and compare them to the
equilibrium distribution at the bath temperature (dashed line). Below (above) the critical wavenumbers $k_\mathrm{crit}$ ($k_L-k_\mathrm{crit}$) the absorption or emission of bath excitations  is strongly suppressed, since in an infinite system it would not be possible to conserve both energy and momentum in such a process. Namely, for a transition $k\to k'$, the absolute value of the quasiparticle momentum obeys $q>|q_x|=|k-k'|$ (cf.~Fig.~\ref{fig:dyn-occ}(c)) \footnote{Umklapp scattering is included in the argument because the dispersion $\varepsilon_{k'}$ is unchanged under $k' \rightarrow k'+k_L$}, corresponding to quasiparticle energies $E_\mathrm{B}(q)\ge E_\mathrm{B}(|k-k'|)$ that have to match $|\varepsilon_k-\varepsilon_{k'}|$, which is impossible for too small or too large $k'$.
This is illustrated in Fig.~\ref{fig:dyn-occ}(c): Only after shifting the Bogoliubov dispersion $E_\mathrm{B}(q)$ (green line) from the origin to the point
$(k_{\mathrm{crit}}, \varepsilon_{k_{\mathrm{crit}}})$ (dashed-dotted line) or further to $k'\geq k_{\mathrm{crit}}$, there is more than one intersection with the lattice dispersion of the system $\varepsilon_k$ (red line) where for the corresponding $k', k$ energy and momentum conservation can be fulfilled.
Similar behavior has been found for a free impurity immersed in a superfluid \cite{LauschEtAl18} (where such a critical momentum only exists for $m_\mathrm{S}>m_\mathrm{B}$).
Since for the finite lattice of $M$ sites momentum is conserved only approximately, ultimately, for $t\gg t_\text{ref}$, the system thermalizes with temperature $T$. 

\begin{figure}
	\centering
	\includegraphics[scale=0.95]{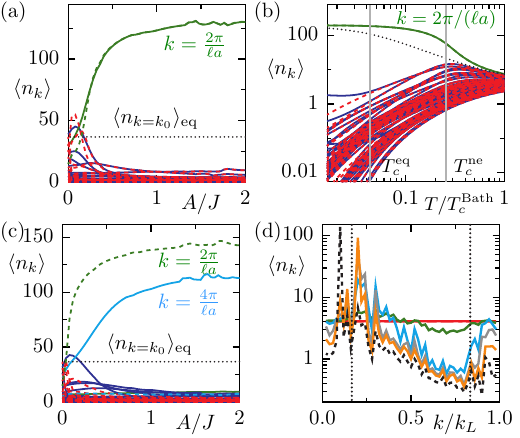}
	\caption[]{Mean-field solutions for mean occupations $\langle n_k\rangle$ of quasimomenta $k$ versus (a) driving strength $A$  and (b) bath temperature $T$ for $t\to \infty$ (dashed lines) and $t=t_\text{ref}$ (solid lines). Dotted line: thermal ground-state occupation. Parameter as in  Fig.~\ref{fig:main}(c), where  $G=0.05 E_R$.
	(c)~Like (a) but for increased intrabath interations $G=0.1 E_R$. At $t=t_\text{ref}$ a metastable condensate is formed in the mode $k=4\pi/(\ell a)$. (d) Like Fig.~\ref{fig:dyn-occ}(b) but for $G=0.1 E_R$.}
	\label{fig:occ-func-a}
\end{figure}

 \emph{Driven-dissipative system.}--We now turn to the dynamics of the driven system with driving amplitude $A=J$, frequency $\hbar\omega=1.5 J$,  and position $\ell=20$. Unlike in the typical regime of Floquet engineering, the driving frequency is \emph{not} large compared to the system's bandwidth $4J$. Thus, driving-induced heating via resonant excitation \cite{EckardtAnisimovas15} is not suppressed, but a dominant impact of the drive. 
 In Fig.~\ref{fig:main}(c) we depict the evolution of the mean occupations of the Floquet modes $\alpha$ on the same time
 interval as in Fig.~\ref{fig:main}(b).  
 We can observe that, in contrast to the undriven scenario, the system quickly relaxes to a steady state on a time $t\lesssim t_\text{ref}/2$. This is not surprising, since the local driving potential breaks both energy and approximate momentum conservation, which suppress the relaxation in the undriven case. 

However, the most striking effect 
is that in the presence of driving the system approaches a non-equilibrium steady state, where more than 60 percent of the particles form a non-equilibrium Bose condensate by occupying the same single-particle Floquet state. Remarkably, this effect happens despite the Floquet (i.e.\ driving-induced) heating. In fact we will argue below that it is actually the interplay between the driving-induced heating and the dissipation from the bath, which causes the non-equilibrium condensation. In Fig.~\ref{fig:dyn-occ}(b) we plot snapshots of the mean occupations $\langle\hat{n}_k\rangle$ of the undriven eigenmodes and we can see that the condensate occurs in a state having a large overlap with the excited undriven eigenstate $\ket{\psi_{k_c}}$ with wavenumber $k_c=2\pi/(\ell a)$. Plotting the occupation of this mode as blue bullets in Fig.~\ref{fig:main}(c), we can see that it perfectly matches the occupation of the most populated Floquet mode, indicating that both modes are in fact identical. This leads us to the following intuitive explanation of the at first glance counterintuitive effect. 

The mode $k_c$ possesses a node at lattice site $\ell$. It is the lowest-energy state of a series of modes $k_\nu=\nu 2\pi/(\ell a)$, with $\nu=1,2,\ldots$, which decouple from the drive,  since $\sin(k_\nu a \ell)=0$. These undriven eigenstates remain Floquet states of the driven system, as can be inferred from Fig.~\ref{fig:dyn-occ}(d), where we plot for each undriven mode $k$, its maximum overlap $\max_\alpha |\langle \psi_k|u_\alpha(0)\rangle|^2$ with a Floquet mode $\alpha$. We can clearly see that the $k_\nu$, which are colored green, correspond to Floquet modes of the problem. In Fig.~\ref{fig:dyn-occ}(e) we plot the (period-averaged) standard deviation of the energy of each Floquet mode $\alpha$. As a result of the strong and resonant driving, almost all modes are well delocalized in energy, except for those Floquet states corresponding to the undriven modes $k_\nu$ (green bullets), which possess sharp energies. We can conclude that the undriven modes $k_\nu$ play a role very similar to quantum many-body-scar states  \cite{Serbyn21, Moudgalya22, Chandran22}. In a quantum many-body system, where the majority of states follows the eigenstate thermalization hypothesis (ETH), so that they give rise to thermal expectation values, such many-body scars correspond to a small set of non-ergodic quantum states violating the ETH. In a many-body Floquet system, a generic Floquet state is completely delocalized in energy, corresponding to an infinite-temperature state \cite{DAlessioRigol14,LazaridesEtAl14b}. The Floquet modes $|u_\alpha(t)\rangle$ of our system are single-particle states of a finite system and, thus, cannot be expected to provide infinite-temperature expectation values. Nevertheless, we can clearly see that they are broad in energy, while, in contrast, the undriven modes $k_\nu$ have sharp energies and, in this sense, resemble quantum many-body scars \cite{Scars}. 

We can now divide the system into two parts, given by the minority of undriven scar-like modes $k_\nu$, which are local and well separated in energy, on the one hand, and the majority of ``actual'' Floquet states on the other. As a consequence of the large level splitting, within the undriven subspace the bath will efficiently transport particles to the lowest state, $k_c$. Moreover, among the driven states the mixing of low-energy with more excited states, raises their energies on average,  corresponding to Floquet heating, so that, as a result, we find also an enhanced bath-induced transfer from driven modes towards $k_c$ (this might be viewed as a Floquet-heating-induced thermoelectric effect). Both effects together, then lead to the observed non-equilibrium Bose condensation in the excited state $k_c$. 

A somewhat related, but in its origin rather different effect can be observed, when instead of being driven periodically, the system is coupled to a very hot local bath \cite{SchnellEtAl17}. However, in stark contrast to the driven system, in this scenario the eigenstate structure of the system is not altered at all. Moreover, the experimental realization a local drive is much easier than that of an additional hot local bath.

The non-equilibrium condensation does not rely on fine tuning. We equally find it for Caesium atoms in a Rubidium bath \cite{SM}. In Fig.~\ref{fig:occ-func-a} we, also, investigate its dependence on both the driving strength $A/J$ (a) and the bath temperature $T/T_c^\text{bath}$ (b) by plotting $\langle \hat{n}_k\rangle$ at time $t/t_\text{ref}=1$ (solid line) and in the steady state (dashed lines) (the dotted line is the equilibrium ground-state occupation).  The majority of particles occupies a single mode for $A/J\gtrsim 0.5$ (a) and $T\lesssim T_c^\text{ne}=JN/M \approx 4 J\gg T_c^\text{eq}\approx0.69J$ (b). Here $T_c^\text{ne}$ denotes a rough estimate of the non-equilibrium condensation temperature below which half the particles occupy $k_c$. It is obtained by assuming that within the cold undriven modes the large level splitting causes the bath to transport all particles to $k_c$, while the remaining $N'$ particles are equally distributed over the driven modes. $T_c^\text{ne}$ then results from setting $N'=N/2$ and approximating the coupling between driven and undriven modes by non-resonant ($K=0$) transitions~\cite{SM}.

\emph{Transient condensation.}---
Also for the driven system the relaxation is slower for $k<k_\text{crit}$. 
When $k_c<k_\text{crit}< k_2$ (which can be achieved by increasing $G/E_R$ to $0.1$, e.g., via the bath density $n_\mathrm{B}$), we observe that during the evolution first the undriven mode $k_2=2k_c=4\pi/(\ell a)$ acquires a large occupation [see Figs.~\ref{fig:occ-func-a}(c) and (d)]. However, ultimately the system relaxes to its true steady state with a condensate in $k=k_c$ (dashed lines). 


\emph{Conclusions.}--- 
We have described a new mechanism for the controlled preparation of non-equilibrium steady states. It relies on the combination of bath-induced dissipation and the driving-induced engineering of scar-like quantum modes that, unlike all other Floquet modes, do not suffer from Floquet heating. Using realistic parameters for a system of bosonic atoms in an optical lattice, we have demonstrated the preparation of an excited-state Bose condensate for bath temperatures well above the equilibrium condensation temperature. It will be interesting to explore the construction of similar protocols for the preparation of correlated target states of interacting systems. 

\begin{acknowledgments}
This work was supported by the Deutsche Forschungsgemeinschaft (DFG, German Research Foundation) via the Collaborative Research Centers SFB/TR185 (Project No.\ 277625399) and SFB 910 (Project No.\ 163436311).
\end{acknowledgments}

\bibliography{mybib,BibOned,BibBasics,BibNumCondensates}

\begin{thebibliography}{82}
\bibitem [{Sca()}]{Scars}%
  \BibitemOpen
  \href@noop {} {}\bibinfo {note} {We avoid the term ``quantum scar'' without
  the attribute ``many-body'', since this term was coined specifically for
  quantum states related to unstable periodic orbits in a corresponding
  classical model.}\BibitemShut {Stop}%
\end{thebibliography}

\end{document}


\title{Supplemental material for ``Floquet-heating-induced Bose condensation in a scar-like mode of an open driven optical-lattice system''}
\author{Alexander~Schnell} 
\email[Electronic address: ]{schnell@tu-berlin.de}
\affiliation{Technische Universit{\"a}t Berlin, Institut f{\"u}r Theoretische Physik, 10623 Berlin, Germany}
\author{Ling-Na~Wu} 
\affiliation{Technische Universit{\"a}t Berlin, Institut f{\"u}r Theoretische Physik, 10623 Berlin, Germany}
\author{Artur~Widera} 
\affiliation{Department of Physics and State Research Center OPTIMAS, Technische Universität Kaiserslautern, 67663 Kaiserslautern, Germany}
\author{Andr{\'e}~Eckardt} 
\email[Electronic address: ]{eckardt@tu-berlin.de}
\affiliation{Technische Universit{\"a}t Berlin, Institut f{\"u}r Theoretische Physik, 10623 Berlin, Germany}

\date{\today}

\maketitle

\section{Bath Hamiltonian and system-bath coupling}
The bath is given by a system of weakly interacting bosonic atoms described by the Hamiltonian 
\begin{equation}
	\hat H_\mathrm{B}=\int_{r} \left\lbrace \hat \chi^\dagger(\vec{r}) \left[\frac{- \hbar^2}{2m_\mathrm{B}} \vec{\nabla}^2\right] \hat \chi(\vec{r}) + \frac{g}{2}\hat \chi^\dagger(\vec{r}) \hat \chi^\dagger(\vec{r}) \hat \chi(\vec{r}) \hat \chi(\vec{r}) \right\rbrace, 
	\label{eq:H_B}
\end{equation}
where we use the convention $\int_r = \intd{^3r}$.  Moreover, $\hat \chi(\vec{r})$ denotes the bosonic field operator, $m_\text{B}$ the mass, and $g$ the contact interaction strength of the bath particles. To find an effective low-energy and low-temperature description of the bath Hamiltonian, we perform the usual  Bogoliubov approximation.  We assume that the extent of the bath is large compared to the system, so that the bath's density $n_\text{B}$ is approximately homogeneous, where it overlaps with the system. Its bulk properties are thus approximated by assuming a homogenous system of $N_\mathrm{B}$ particles in a volume $V$ with periodic boundary conditions, so that $n_\text{B}=N_\text{B}/V$. After defining the annihilation operators for a bath particle of momentum $\vec{q}$ (we reserve the symbol $\vec{q}$ for bath momenta as a convention), $\hat c_{\vec{q}} = \frac{1}{\sqrt{V}} \int_{r}  \mathrm{e}^{-i \vec{q}\vec{r}} \hat \chi(\vec{r})$, for temperatures $T$ well below the critical bath temperature $T^\mathrm{Bath}_c= {2\pi \hbar^2}({ n_\mathrm{B}}/{\zeta({3/2})})^{2/3} /{\kboltz m_\mathrm{B}}$ (we set $k_\mathrm{B}=1$ throughout the manuscript and neglect the small change of $T^\mathrm{Bath}_c$ due to the presence of interactions)  and weak interactions $g\ll 4\pi\hbar^2(V/N_\mathrm{B})^{1/3}/m_\mathrm{B}$ one may represent the field operator $\hat \chi$ as $\hat \chi(\vec{r}) =\chi_0 + \delta \hat \chi(\vec{r})$ with c-number field  $\chi_0 = \sqrt{N_0/V}$ (with ground state occupation $N_0=\langle \hat c_0^\dagger \hat c_0 \rangle$) describing the condensate, and small operator-valued fluctuations 
\begin{align}
\delta\hat \chi(\vec{r})= \frac{1}{\sqrt{V}} \sum_{\vec{q}\neq0} \mathrm{e}^{i \vec{q}\vec{r}} \hat c_{\vec{q}}.
\label{eq:def-delxi}
\end{align} 
Plugging the decomposition into the bath Hamiltonian, Eq.~\eqref{eq:H_B} and keeping fluctations only up to quadratic order in 
$\delta \hat \chi(\vec{r})$, we obtain
\begin{align}
	\begin{split}
	\hat H_\mathrm{B}&\approx\int_{r}  \delta \hat \chi^\dagger(\vec{r}) \left[- \frac{\hbar^2}{2m_\mathrm{B}} \vec{\nabla}^2 + G\right] \delta\hat \chi(\vec{r}) + \frac{GN_\mathrm{B}}{2}  \\
	&+ \frac{G}{2}\int_{r}  \left[ \delta \hat \chi(\vec{r}) \delta \hat \chi(\vec{r}) + \delta\hat \chi^\dagger(\vec{r}) \delta \hat\chi^\dagger(\vec{r}) \right].
	\label{eq:H_B-delta-2}
	\end{split}
\end{align}
Here we have used $N_B= N_0+\int \delta \hat \chi^\dag(\vec{r})\delta \hat \chi(\vec{r})$ and introduced $G=gn_\mathrm{B}$.
We then use Eq.~\eqref{eq:def-delxi} and perform the standard Bogoliubov transformation 
\begin{align}
\hat \beta_{\vec{q}} = u_q \hat c_{\vec{q}} + v_q \hat c^\dagger_{-\vec{q}}
\end{align}
to bring the Hamiltonian to the form 
\begin{align}
\hat H_\mathrm{B}=\sum_{\vec{q}}  E_\mathrm{B}(q) \hat\beta^\dagger_{\vec{q}}\hat\beta_{\vec{q}},
\end{align}
with Bogoliubov dispersion 
\begin{align}
E_\mathrm{B}(q) =  \sqrt{E_0(q)^2+2G E_0(q)},
\label{eq:Bog-disp}
\end{align}
where $E_0(q)=\hbar^2q^2/(2m_\mathrm{B})$, 
and the transformation follows from $u_q^2-v_q^2=1$ and $u_q v_q = G/(2 E_\mathrm{B}(q))$.

The system-bath coupling Hamiltonian
\begin{align}
 	\hat{H}_\mathrm{SB}(t) =\gamma \int_{r} \hat \Psi^\dagger(\vec r, t) \hat \Psi(\vec r, t)  \hat B(\vec{r})
\end{align}
can be expressed in terms of Bogoliubov quasiparticles
\begin{align}
	 & \hat B(\vec{r})=  \hat\chi^\dagger(\vec{r})   \hat \chi(\vec{r}) - n_\mathrm{B}\\
&=  \sqrt{{n_\mathrm{B}}} \left[ \delta \hat \chi(\vec{r}) + \delta \hat \chi^\dagger(\vec{r}) \right] +\mathcal{O}(\delta\hat \chi^2)\\
&=  \sqrt{\frac{n_\mathrm{B}}{V}} \sum_{\vec{q}\neq0} (u_q-v_q)\left[\mathrm{e}^{i \vec{q}\vec{r}} \hat \beta_{\vec{q}} + \mathrm{e}^{-i \vec{q}\vec{r}} \hat \beta_{\vec{q}}^\dagger  \right] +\mathcal{O}(\delta\hat \chi^2).
\end{align}
In the last step we have employed Eq.~\eqref{eq:def-delxi} as well as the inverse Bogoliubov transformation $\hat c_{\vec{q}}=u_q \hat{\beta}_{\vec{q}} - v_q  \hat{\beta}_{-\vec{q}}^\dagger$.

Within the tight-binding approximation, the field operator of the system is expanded in terms of the lowest-band Wannier states $| i\rangle$,  $\hat \Psi(\vec r, t) = \sum_{i=1}^{M} w_i(\vec{r}) \hat a_i$, with Wannier functions $w_i(\vec{r})=\langle\vec{r}|i\rangle$ and corresponding annihilation operators $\hat a_i$. Thus, in leading order in $\delta\hat \chi$ we have
\begin{align}
 	\hat{H}_\mathrm{SB}(t) = \gamma \sum_{i,j, \vec{q}\neq0}  \hat a_i^\dagger \hat a_j \left[\kappa_{ij}(q)\hat \beta_{\vec{q}} + \kappa_{ji}(q)^* \hat \beta_{\vec{q}}^\dagger  \right], 
\end{align}
with coefficients 
\begin{align}
 	\kappa_{ij}(q)= \sqrt{\frac{n_\mathrm{B} E_0(q)}{V E_\mathrm{B}(q)}}  \int_{r}  w_i(\vec{r})^* w_j(\vec{r}) \mathrm{e}^{i \vec{q}\vec{r}} \approx \delta_{ij}  \kappa_{i}(q).
\end{align}
In the last step we neglect all contributions from off-site Wannier orbitals $w_i(\vec{r})^*w_j(\vec{r}) \approx \delta_{ij} \vert w_i(\vec{r})\vert^2$.
Thus, the system--bath coupling operator, is brought to the standard form of a Hubbard-Holstein model  \cite{KleinEtAl07,BrudererEtAl07}
\begin{align}
 	\hat{H}_\mathrm{SB}(t) = \gamma \sum_i \hat  n_i \sum_{\vec{q}\neq0} \left[\kappa_{i}(q)\hat \beta_{\vec{q}} + \kappa_{i}(q)^* \hat \beta_{\vec{q}}^\dagger  \right]\equiv  \gamma \sum_{i} \hat  n_i \hat B_i.
\end{align}
The system coupling operators $\hat  n_i= \hat a_i^\dagger \hat a_i$ couple to the phononic Bogoliubov modes through bath operators $\hat B_i$ at each individual site $i$. 

In order to obtain a simple analytical expression for the coefficients $\kappa_{i}(q)$, we approximate the Wannier functions by harmonic oscillator ground states 
\begin{align}
	w_i(\vec{r}) \approx \varphi^\mathrm{HO}_L(x-x_i) \varphi^\mathrm{HO}_T(y) \varphi^\mathrm{HO}_T(z)
\end{align}
with site position $x_i = i a$. With the oscillator frequency in the lattice minimum $\Omega_L=2\sqrt{V_0 E_R}/\hbar$, one has
 $\varphi^\mathrm{HO}_L(x) 
  =( d_L \sqrt{\pi})^{-0.5} \mathrm{e}^{-\left({x}/{d_L}\right)^2/2}$ with oscillator length
 $d_L=\sqrt{\hbar/m\Omega_L}$. In transverse direction  $\varphi^\mathrm{HO}_T$ is defined equivalently, with oscillator length $d_T$. This yields
\begin{align}
 	\kappa_{i}(q)= \sqrt{\frac{n_\mathrm{B} E_0(q)}{V E_\mathrm{B}(q)}}   \mathrm{e}^{i {q_x}x_i} \mathrm{e}^{-\frac{1}{4}\left[{d_L^2}{q_x^2}+{d_T^2}({q_y^2+q_z^2})\right]}.
\end{align}


\section{Floquet-Born-Markov-secular approximation}
In the system-bath coupling Hamiltonian, we have omitted contributions beyond the linear order $\delta \hat \chi(\vec{r})$. 
This means that we restrict ourselves to one-phonon scattering in the bath, which largely dominates over higher-order phonon scattering for low temperatures $T$ \cite{LauschEtAl18}. As a consequence  $\hat H_\mathrm{SB}$ is already in the form $\hat H_\mathrm{SB}=\sum_{i} \hat v_i \otimes \hat B_i$
required for the Floquet-Born-Markov formalism \cite{BreuerPetruccione,KohlerEtAl97,BreuerEtAl00,HoneEtAl09,Wustmann10}. Here the part of the coupling operator acting in the system's state space is given by $\hat{v}_i=\hat{n}_i$. Since we are dealing with non-interacing particles in the system, we can obtain the many-particle master equation from the single-particle one. For the single-particle problem, we have to replace $\hat{n}_i$ by $|i\rangle\langle i|$, giving $\hat{v}_i=|i\rangle\langle i|$. In the limit of weak system--bath coupling, where the rotating wave (or secular) approximation is valid, one finds golden rule-type rates \cite{BreuerPetruccione,KohlerEtAl97,BreuerEtAl00,HoneEtAl09,Wustmann10,DiermannEtAl19}
\begin{align}
	R_{\alpha \beta} = \frac{2\pi \gamma^2}{\hbar} \mathrm{Re} \sum_{K\in \mathbb{Z}} \sum_{ij} (v_i)^{(K)*}_{\alpha \beta}(v_j)^{(K)}_{\alpha \beta} W_{ij}(\Delta^{(K)}_{\alpha \beta}),
	\label{eq:rates-general}
\end{align}
for a bath-induced quantum jump of a single particle from Floquet state $\beta$ to Floquet state $\alpha$. 
Here we have defined the quasienergy difference $\Delta^{(K)}_{\alpha \beta} =\varepsilon_{\alpha}-\varepsilon_\beta+K\hbar \omega$, and the Fourier components of the coupling matrix elements
\begin{align}
(v_i)^{(K)}_{\alpha \beta}&= \frac{1}{\mathcal{T}} \int_0^{\mathcal{T}} \mathrm{d}t \mathrm{e}^{-iK\omega t} \braket{u_\alpha(t)}{i} \braket{i}{u_\beta(t)},
 	\label{eq:v-floq}\\
	&=\sum_{r}  u^{(r)*}_{\alpha,i} u^{(r+K)}_{\beta,i} .
 \end{align}
Here $\mathcal{T}=2\pi/\omega$ is the driving period and 
 $\ket{u_\alpha(t)} $ a Floquet mode, the $r$-th Fourier component  of which is denoted by $u^{(r)}_{\alpha,i}= \braket{i}{u_{\alpha}}^{(r)}$. 
We have also employed the half-sided Fourier transform 
\begin{align}
W_{ij}(E) = \frac{1}{\pi \hbar} \int_0^\infty \! \mathrm{d}\tau \mathrm{e}^{-\frac{i}{\hbar}E\tau} \langle {\tilde{B}}_i(\tau) \hat B_j \rangle_\mathrm{B}
 \end{align} of the bath correlation function. Here $\langle \cdot \rangle_\mathrm{B} = \mathrm{Tr}_\mathrm{B} ( \hat  \varrho_\mathrm{B} \cdot)$ and where $\tilde{O}(\tau)$ indicates the operator $\hat{O}$ in the interaction picture, 
  \begin{align}
\tilde{O}(\tau)  = \mathrm{e}^{i(\hat H_\mathrm{S} + \hat H_\mathrm{B})\tau} \hat O \mathrm{e}^{-i(\hat H_\mathrm{S} +\hat H_\mathrm{B})\tau}.
  \end{align}

We use that the bath is in a thermal state $\hat \varrho_\mathrm{B}=  \frac{1}{Z} \exp(-\hat H_\mathrm{B}/\kboltz T)$, to  evaluate
\begin{align}
 \langle \tilde{B}_i(t)& \hat  B_j \rangle_\mathrm{B} \\
 \begin{split}
 = \sum_{\vec{q},\vec{q}^{\prime}\neq 0}\!\! & \!\left\langle \!\left[\kappa_{i}(q)\hat \beta_{\vec{q}} \, \mathrm{e}^{-\frac{i}{\hbar}E_\mathrm{B}(q) t} + \kappa_{i}(q)^* \hat \beta_{\vec{q}}^\dagger \,  \mathrm{e}^{\frac{i}{\hbar}E_\mathrm{B}(q) t} \right]\! \right. \\ 
& \qquad \left. \left[\kappa_{j}(q^{\prime})\hat \beta_{\vec{q}^{\prime}} + \kappa_{j}(q^{\prime})^* \hat \beta_{\vec{q}^{\prime}}^\dagger  \right]\!\right\rangle_\mathrm{B}
\end{split}\\
= \int_{-\infty}^\infty\! &\mathrm{d}E \ J_{ij}(E) \mathrm{e}^{\frac{i}{\hbar}E t} n(E)
\end{align}
with  Bose-Einstein occupation function  
 \begin{align}
 	n(E) = \frac{1}{\mathrm{e}^{E/\kboltz  T}-1},
 \end{align}
and spectral density
 \begin{align}
 \begin{split}
 	J_{ij}(E) = \sum_{\vec{q}\neq 0}& \left[ \kappa_{i}(q)^*\kappa_{j}(q) \delta(E-E_\mathrm{B}(q))\right.\\
				& \left. -\kappa_{i}(q)\kappa_{j}(q)^* \delta(E+E_\mathrm{B}(q))\right].
\end{split}
 \end{align}
Therefore, using the Sokhotski–Plemelj formula and neglecting the imaginary part of $W_{ij}$ (giving rise to a lamb shift, which in the secular coupling limit becomes negligible), we have
\begin{align}
	W_{ij}(E) = J_{ij}(E)  n(E).
\end{align}

Finally, we take the continuum limit for the bath sum over~$\vec{q}$, 
$\frac{(2\pi)^3}{V} \sum_{\vec{q}} \rightarrow \int\mathrm{d}^3 q$,
 \begin{align}
 \begin{split}
J_{ij}(E) \overset{E>0}{=}  & \frac{n_\mathrm{B}}{(2\pi)^3}\!\int\!\mathrm{d}^3 q \frac{E_0(q)}{E_\mathrm{B}(q)}\\
&\times \mathrm{e}^{-\frac{1}{2}\left[{d_L^2}{q_x^2}+{d_T^2}({q_y^2+q_z^2})\right]}\mathrm{e}^{i {q_x}(x_j - x_i)} \delta(E-E_\mathrm{B}(q)),
\end{split}
 \end{align}
  and $J_{ij}(-E) {=}- J_{ij}(E)^*$. 
We introduce spherical coordinates with the  $x$-axis being the polar axis and set $\zeta=\cos\theta$, to find
 \begin{align}
J_{ij}(E) \overset{E>0}{=}  
 \frac{n_\mathrm{B}}{(2\pi)^2}\int_0^\infty\!\!\mathrm{d} q &  q^2 \frac{E_0(q)}{E_\mathrm{B}(q)}I_{ij}(q)\delta(E-E_\mathrm{B}(q)),
 \end{align}
 with function
  \begin{align}
I_{ij}(q) {=}
 \int_{-1}^1\!\!\mathrm{d} \zeta\mathrm{e}^{-\frac{q^2}{2}\left[{d_L^2}{\zeta^2}+{d_T^2}(1-\zeta^2)\right]}  \mathrm{e}^{i {q} \zeta(x_j - x_i)}=I_{ij}(q)^*.
\label{eq:I-integral-2}
 \end{align}
We solve Eq.~\eqref{eq:Bog-disp} for the momentum
 \begin{align}
 q(E)= \frac{\sqrt{2m_\mathrm{B}}}{\hbar} \left(\sqrt{E^2+G^2}-G\right)^{1/2}
  \end{align}
 of a Bogoliubov quasiparticle at energy $E$. This allows us to transform the differential
   \begin{align}
 2q \mathrm{d}q= \frac{{2m_\mathrm{B}}}{\hbar^2} \frac{E_\mathrm{B}}{\sqrt{E_\mathrm{B}^2+G^2}} \mathrm{d}E_\mathrm{B}.
  \end{align}
  After transforming the $q$-integral into an integral over $E_\mathrm{B}$, we can directly evaluate the delta distribution and find
   \begin{align}
J_{ij}(E) &{=} \mathrm{sgn}(E) 
 \frac{n_\mathrm{B}}{(2\pi)^2 2}\frac{q(E)^3}{\sqrt{E^2+G^2}}I_{ij}(q(E)).
 \end{align}
  Note that for small energies $E \ll G$ we find super-ohmic behavior $J(E)\propto E^{3}$, 
  while for $E \gg G$ the spectral density decays again, due to the exponential decay of $I_{ij}(q(E))$.

It is left to evaluate the function $I_{ij}(q)$.  Since the integral over $\zeta$ is hard to evaluate in general, we restrict us for practical reasons to the case where $d_L=d_T$, so that  we find 
 \begin{align}
I_{ij}(q)= \mathrm{e}^{-\frac{1}{2} q^2 d_T^2} 2 \mathrm{sinc}\!\left[q(x_i-x_j)\right], 
  \end{align}
with $\mathrm{sinc}(x) = \sin(x)/ x$.
 Note that we expect similar results for the dynamics also in the general case where $d_L \ne d_T$ as long as all $q$ fulfill $q\ll \vert d_L^2-d_T^2\vert^{-1/2}$. This can be seen from approximating $\mathrm{e}^{-\frac{1}{2} q^2 (d_L^2-d_T^2)}\approx1$ in the integral of  Eq.~\eqref{eq:I-integral-2}, which is a good approximation for such values $q$.
 
%

\section{Condensation temperature in equilibrium}
\begin{figure}
	\includegraphics[scale=1.2]{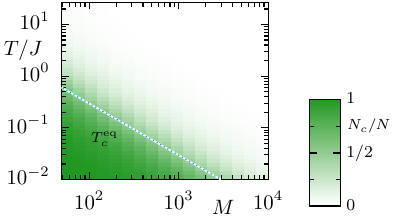}
	\caption{Condensate fraction $N_c/N$ for a tight binding chain of $M$ sites 
    at temperature $T$ (shading). The blue-white dotted line gives the 
    analytical estimate, Eq.~\eqref{eq:T_ceq}, for the condensation temperature, where half of the 
    particles occupy the single-particle ground state.}
	\label{fig:equilibrium}    
\end{figure}
These results have already been presented in the supplemental material of Ref.~\cite{SchnellEtAl17}, 
but are included here again for completeness.
Under equilibrium conditions, when the system is coupled only to the bath 
of temperature $T$ (i.e.\ for $A=0$), the mean-field equation (see main text) is 
solved by the grand-canonical mean occupations
\be
\la\no_k\ra = \frac{1}{e^{ (\varepsilon_k-\mu)/T}-1}
\ee
with chemical potential $\mu$. 
When (finite-size) Bose condensation sets in, $\mu$ approaches $\varepsilon_{k_0}$ 
from below, so that the occupations of the low-energy modes with $k\ll1$ can be 
approximated by
\be
\la\no_k\ra \simeq  \frac{T}{\varepsilon_k-\mu} 
            \simeq \frac{T}{J k^2a^2 -2J -\mu},
\ee
where we have used $\varepsilon_k =-2J\cos(ka)\simeq -2J+Jk^2a^2$. The chemical 
potential can be expressed in terms of the occupation $N_c=\la \no_{k_0}\ra$ of 
the ground state with wave number $k_0=\pi/[a(M+1)]$, 
\be
\mu= -2J + Jk_0^2a^2 -T/N_c.
\ee
For low temperatures, the number $N'$ of particles occupying excited states, 
with $k=\nu\pi/[a(M+1)]$, is dominated by the long-wavelength modes, so that we can 
approximate
\be
N' = \sum_{k'\ne k_0} \la\no_k\ra
   \simeq \sum_{\nu=2}^\infty 
        \frac{1}{\frac{J\pi^2}{TM^2}(\nu^2-1)+\frac{1}{N_c}}.
\ee
For a finite system, we define the characteristic temperature $T_c$, where Bose 
condensation sets in, as the temperature for which half of the particles occupy 
the single-particle ground state, $N'=N_c=N/2$. It is given
by 
\be\label{eq:T_ceq}
T^\text{eq}_c \simeq \frac{c\pi^2}{2} \frac{nJ}{M} \approx  8.3\,\frac{nJ}{M},
\ee
where $n=N/M$, and   $c\approx1.68$ solves $1=c\sum_{\nu=2}^\infty 1/(\nu^2 + c-1)$. In
Fig.~\ref{fig:equilibrium} we plot the ground-state occupation (i.e.\
the condensate fraction) of the tight binding chain together with the estimate
(\ref{eq:T_ceq}) for the condensation temperature $T^\text{eq}_c$. The inverse 
dependence of $T^\text{eq}_c$ on the system size $M$ reflects the well-known 
result that in one spatial dimension, in the thermodynamic limit Bose-Einstein 
condensation is suppressed by thermal long-wavelength fluctuations.

\section{Nonequilibrium condensation temperature}

Here we present a very rough estimate for the characteristic temperature $T_{c}^\mathrm{ne}$ for nonequilibrium
condensation. 
As discussed in the main text, we can distinguish modes that decouple from the drive, which form the scar-like part (or cold part) of the system, and the remaining hot part of the system, which is subjected to strong resonant driving. Since the cold modes are few and equally spaced in momentum, they are well separated in energy. As a result, the bath transfers essentially all particles within the cold part to the cold mode of lowest energy, $k_0$. Within the hot part of the system, the driving mixes states of different energy, so that roughly all Floquet modes acquire the same occupation. 
This suggests the following approximation for the occupation numbers:
\begin{equation}
	\langle \hat n_\alpha \rangle = \left\lbrace \begin{array}{cc}
	N_c & \text{for } \alpha = \alpha_c\\
	N'/M & \text{else,}
	\end{array}
	 \right.
\end{equation}
where $N=N_c+N'$ (cf.~Fig.~2(b) in the main text). Here we have neglected the excited cold modes, which are few and whose occupations are small. 
In the condensate regime,  $N_c \sim N$, one has $N_c \gg N'/M$, and hence the dominating terms $\propto N_c$ in the mean-field equation (see main text) read for $\alpha = \alpha_c$
\begin{equation}
	0 = N_c \sum_{\beta \neq \alpha_c} \left(A_{\alpha_c \beta} \frac{N'}{M} - R_{\beta \alpha_c} \right) + \mathcal{O}(N_c^0),
	\label{eq:Tcne-eom}
\end{equation}
with $A_{\alpha\beta} =R_{\alpha\beta}-R_{\beta \alpha}$. 
Since the factor $I_{ij}(q)$, which enters in the rates $R_{\beta \alpha_c}$, has its dominating contribution for $i=j$, we may approximate 
\begin{equation}
R_{\alpha\beta} \approx \frac{2\pi \gamma^2}{\hbar} \sum_{K\in \mathbb{Z}} \sum_{i} \vert (v_i)^{(K)}_{\alpha \beta} \vert^2 \frac{J(\Delta^{(K)}_{\alpha \beta})}{{e}^{ \Delta^{(K)}_{\alpha \beta}/T}-1}.
\end{equation}
with $J(E)=J_{ii}(E)$ which is independent of $i$.
This gives 
\begin{equation}
A_{\alpha\beta} \approx \frac{2\pi \gamma^2}{\hbar} \sum_{K\in \mathbb{Z}} \sum_{i} \vert (v_i)^{(K)}_{\alpha \beta} \vert^2 [- J(\Delta^{(K)}_{\alpha \beta})],
\end{equation}
In Eq.~\eqref{eq:Tcne-eom} we divide by $(2\pi \gamma^2N_c)/\hbar$ and have
\begin{equation}
0 =  \sum_{\beta \neq \alpha_c} \sum_{K\in \mathbb{Z}} \sum_{i} \vert (v_i)^{(K)}_{ \beta\alpha_c} \vert^2 J(\Delta^{(K)}_{ \beta\alpha_c}) \left(\frac{N'}{M} - \frac{1}{{e}^{ \Delta^{(K)}_{\beta\alpha_c}/T}-1}\right).
\end{equation}
Here, due to the strong driving $A$ and the low frequency $\omega$, the matrix elements $(v_i)^{(K)}_{ \beta\alpha_c}$ and quasienergy differences $\Delta^{(K)}_{\beta\alpha_c}$ typically can only be determined numerically. 

Nevertheless a very rough estimate can be found by requiring that the term in the brackets is zero after averaging over all states, and that the
most prominent contribution stems from photon index $K=0$, 
\begin{equation}
0 =  \left\langle \frac{N'}{M} - \frac{1}{{e}^{ \Delta^{(0)}_{\beta\alpha_c}/T}-1}\right\rangle_\beta \approx   \frac{N'}{M} - \frac{1}{{e}^{\left\langle \Delta^{(0)}_{\beta\alpha_c}\right\rangle_\beta/T}-1} .
\end{equation}
We use the convention to choose the quasienergies (which are defined modulo $\hbar\omega$ only) so that they approach the energy eigenvalues in the limit of vanishing drive ($A=0$).
Assuming that $\left\langle \Delta^{(0)}_{\beta\alpha_c}\right\rangle_\beta \approx 2J$, which corresponds to the value without the driving $A=0$,
and defining the condensation temperature $T_c^\text{ne}$ by the condition that $N_c=N/2$ (and thus $N'=N/2$), we find
\begin{equation}
T_c^\mathrm{ne}=  \frac{2J}{\log(2/n+1)} \overset{n\gtrsim 1}{\approx} Jn,
\end{equation}
with filling factor $n=N/M$.

\hide{Here we assume that the Floquet states are (inspite of the driving) still roughly given by momentum eigenstates $\alpha = k$, with $k = \nu \pi/[(M+1)a]$ and $\nu=1, \dots, M$. The dispersion therefore reads 
$E(k) = -2J \cos(ka)$.
Since the involved sum over the states has no IR divergencies, we may approximate it by performing the continuum limit $\sum_{\beta \neq \alpha_c}  = \sum_{k \neq k_c} \rightarrow \frac{a(M+1)}{\pi}  \int_0^{\pi/a} \mathrm{d}k$. 
Finally, we note that without the driving, $A=0$, the factor $\sum_{i} \vert (v_i)^{(K)}_{ \beta\alpha_c} \vert^2$ is independent 
of the state index $\beta$. For finite driving, we therefore assume $\sum_{i} \vert (v_i)^{(K)}_{ \beta\alpha_c} \vert^2 \approx f_K$, i.e.~that the coupling matrix element only depends on photon index $K$ and not on the Floquet state involved. We additionally assume that the major contribution stems from the term $K=0$. 
This gives
\begin{equation}
0 = \int_{0}^{\pi/a}\!\!  \mathrm{d}k {J(E(k)-E_c)} \left(\frac{N'}{M} - \frac{1}{{e}^{\frac{E(k)-E_c}{T}}-1}\right) =I_1-I_2.
\label{eq:I1-I2}
\end{equation}
Even though the condensate is found at a finite momentum $k_c \approx 2\pi/(\ell a)$, for a very rough approximation of $T_c^\mathrm{ne}$, we can use $k_c \approx 0$.
Additionally, we observe that for small energies $E \ll G$, the behavior 
\begin{equation}
J(E)= C E^{3}
\label{eq:J-lowen}
\end{equation}
is found, where for the parameters of Fig.~2 in the manuscript $G \approx 0.775 J$. Even though the regime of validity of Eq.~\eqref{eq:J-lowen}
only spans a smaller fraction of the integral, we use it to approximate both integrals to
\begin{align}
I_1 &= (2J)^3 C \int_{0}^{\pi/a}\!\!  \mathrm{d}k {[1-\cos({ka})]^3} \frac{N'}{M} = \frac{(2J)^3 C5\pi N'}{2aM}\\
I_2 &= (2J)^3 C \int_{0}^{\pi/a}\!\!  \mathrm{d}k \frac{[1-\cos({ka})]^3}{{e}^{{2J(1-\cos({ka}))}/{T}}-1} \\
& \overset{T\gtrsim J}{\approx} T (2J)^2 C \int_{0}^{\pi/a}  \mathrm{d}k {[1-\cos({ka})]^2} = \frac{T(2J)^2 C3\pi }{2a}.
\end{align}
Solving Eq.~\eqref{eq:I1-I2} for the populations of the depletion, yields
\begin{align}
\frac{N'}{M}=\frac{3T}{10J}.
\end{align}
At the critical temperature $N'=N_c=N/2$, which gives
\begin{align}
T_c^\mathrm{ne} = \frac{5}{3}n J .
\end{align}}

\section{Caesium in Rubidium}
\begin{figure}
	\centering
	\includegraphics[]{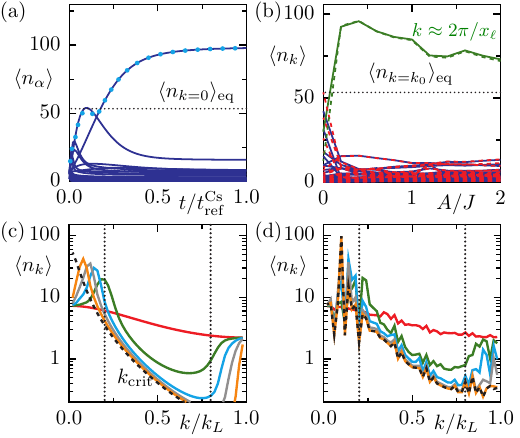}
	\caption[]{Similar plots as in the main text, but for Caesium in a Rubidium bath, $m_\mathrm{B}/m_\mathrm{S}=87/137$, all obtained from mean-field theory. Other parameters: $T=0.1 T_c^{\mathrm{Bath}}, n_\mathrm{B}=1 / a^3, a=100 a_\mathrm{Rb}, \hbar\omega = J$, other parameters as in Fig.~1 and 2 of the main text. $A = 0.3J$ in (a) and (d). Note that we have to choose a reference time scale that is a factor 10 larger than in the main text, $t_\mathrm{ref}^\mathrm{Cs}=160{\pi \hbar^3}/({m_\mathrm{B}k_L n_\mathrm{B}\gamma^2})$, because the condensate lies below $k_\mathrm{crit}$, and we have to wait long enough such that momenta below it relax.}
\label{fig:Caesium}
\end{figure}
When considering Caesium 133 atoms immersed in Rubidium 87, as in the Kaiserslautern experiment \cite{SchmidtEtAl18,SchmidtEtAl19PSS,BoutonEtAl20,SchmidtEtAl19}, we find the same behaviour as discussed for the case of Potassium 39 atoms. This can be seen in Fig.~\ref{fig:Caesium}. However, one difference is that now the time scale for the relaxation now becomes about 10 times larger, since the condensate mode lies below $k_\text{crit}$.  

\bibliography{mybib,BibOned,BibBasics,BibNumCondensates}

\hide{
\section{Single-particle rates for a weakly interacting Bose-condensed Bath}
 Note that $\hat H_\mathrm{SB}$ is already in the desired form $\hat H_\mathrm{SB}=\sum_{i} \hat v_i \otimes \hat B_i$
for the open quantum system formalism if we consider that $\alpha$ is the continuous index $\vec{r}$ and $v(\vec{r}) = \ket{\vec{r}}\bra{\vec{r}}$, $\hat B(\vec{r})= \hat \chi^\dagger(\vec{r})  \hat \chi(\vec{r}) - n_\mathrm{B}$.
The usual Born-, Markov- \cite{BreuerPetruccione} and 
full rotating-wave approximation \cite{BreuerEtAl00,Wustmann10,DiermannEtAl19} lead to single-particle rates
\begin{align}
	R_{\alpha \beta} = \frac{2\pi \gamma^2}{\hbar} \mathrm{Re} \sum_{K\in \mathbb{Z}} \int_{r} \int_{r'} v^{(K)}_{\alpha \beta}(\vec{r})v^{(K)}_{\alpha \beta}(\vec{r}^{\, \prime})^* W(\Delta^{(K)}_{\alpha \beta}, \vec{r}, \vec{r}^{\, \prime}),
	\label{eq:rates-general}
\end{align}
describing a bath-induced quantum jump from Floquet state $\alpha$ to Floquet state $\beta$. 
Here we have defined the quasienergy difference $\Delta^{(K)}_{\alpha \beta} =\varepsilon_{\alpha}-\varepsilon_\beta+K\hbar \omega$, and the Fourier components of the coupling matrix 
\begin{align}
 v^{(K)}_{\alpha \beta}(\vec{r})= \frac{1}{\mathcal{T}} \int_0^{\mathcal{T}} \mathrm{d}t \mathrm{e}^{-iK\omega t} \braket{u_\alpha(t)}{\vec{r}} \braket{\vec{r}}{u_\beta(t)}
 	\label{eq:v-floq}
 \end{align}
with time period $\mathcal{T}=2\pi/\omega$,  
and Floquet mode $\ket{u_\alpha(t)} $.
We have also employed the half-sided Fourier transform 
\begin{align}
W(E,  \vec{r},\vec{r}^{\, \prime}) = \frac{1}{\pi \hbar} \int_0^\infty \! \mathrm{d}\tau e^{-\frac{i}{\hbar}E\tau} \langle {\tilde{B}}(\vec{r},\tau) \hat B(\vec{r}^{\, \prime}) \rangle_\mathrm{B}
 \end{align} of the bath correlation function,  we denote $\langle \cdot \rangle_\mathrm{B} = \mathrm{Tr}_\mathrm{B}  \hat  \varrho_\mathrm{B}   \cdot $ and the tilde in $\tilde{B}(\vec{r},\tau)$ indicates the operator in the interaction picture, generally by the convention
 \begin{align}
\tilde{O}(\tau)  = \mathrm{e}^{i(\hat H_\mathrm{S} + \hat H_\mathrm{B})\tau} \hat O \mathrm{e}^{-i(\hat H_\mathrm{S} +\hat H_\mathrm{B})\tau}.
  \end{align}

In order to treat the bath effectively, we employ the decomposition of the field $\hat \chi(\vec{r}) =\chi_0 + \delta \hat \chi(\vec{r})$
into its constant mean value $\chi_0$ and fluctuations $\delta \hat\chi$ around it.
Plugging in the specific form of 
\begin{align}
	 & \hat B(\vec{r})=  \hat\chi^\dagger(\vec{r})   \hat \chi(\vec{r}) - n_\mathrm{B}\\
	&= \chi^*_0 \chi_0 + \chi^\dagger_0 \delta  \hat\chi(\vec{r}) + \delta  \hat\chi^\dagger(\vec{r})  \chi_0 + \delta \hat\chi^\dagger(\vec{r})  \delta \hat\chi(\vec{r}) -  n_\mathrm{B}\\
&=  \sqrt{{n_\mathrm{B}}} \left[ \delta \hat \chi(\vec{r}) + \delta \hat \chi^\dagger(\vec{r}) \right] +\mathcal{O}(\delta\hat \chi^2).
\end{align}
 into $W(E,  \vec{r},\vec{r}^{\, \prime})$, and again omitting fluctuations of higher order than $\delta \hat \chi(\vec{r})^2$ 
(which means that we restrict ourself to one-phonon scattering in the bath, which largely dominates over higher-order phonon scattering for low temperatures $T$ \cite{LauschEtAl18}), 
we find
\begin{align}
\begin{split}
 &\langle \tilde{B}(\vec{r},t) \hat  B(\vec{r}^{\, \prime}) \rangle_\mathrm{B} \\
 &={n_\mathrm{B}} \left\langle \left( \delta \tilde\chi(\vec{r}, t) + \delta \tilde\chi^\dagger(\vec{r},t)  \right) \left( \delta \hat \chi(\vec{r}^{\, \prime}) + \delta \hat \chi^\dagger(\vec{r}^{\, \prime})  \right)\right\rangle_\mathrm{B}
 \end{split}\\
 &= \frac{{n_\mathrm{B}}}{V}\!\! \sum_{\vec{q},\vec{q}^{\prime}\neq 0}\!\! \mathrm{e}^{i(\vec{q} \vec{r}+\vec{q}^{\prime} \vec{r}^{\prime})}  \!\left\langle \!\left( \tilde{c}_{\vec{q}}(t) + \tilde{c}^\dagger_{-\vec{q}}(t)  \right) \! \left( \hat  c_{\vec{q}^{\, \prime}}+ \hat  {c}^\dagger_{-\vec{q}^{ \prime}} \right)\!\right\rangle_\mathrm{B}.
 \end{align}
In the last step we have used Eq.~\eqref{eq:def-delxi}. We take
$\hat  c_{\vec{q}} + \hat  c^\dagger_{-\vec{q}}= (u_q-v_q) \left(\hat \beta_{\vec{q}} + \hat \beta^\dagger_{-\vec{q}}\right)$
 and
$\tilde{\beta}_{\vec{q}}(t) = \mathrm{e}^{-\frac{i}{\hbar}E_\mathrm{B}(q) t} \hat  \beta_{\vec{q}}$
and that the bath is in a thermal state $\hat \varrho_\mathrm{B}=  \frac{1}{Z} \exp(-\hat H_\mathrm{B}/k_\mathrm{B} T)$, to find
\begin{align}
\begin{split}
& \langle \tilde{B}(\vec{r},t) \hat B(\vec{r}^{\, \prime}) \rangle_\mathrm{B} \\
& = \frac{n_\mathrm{B}}{V} \sum_{\vec{q} \neq 0} {\mathrm{e}^{-i\vec{q} (\vec{r}- \vec{r}^{\prime})} }{(u_q-v_q)^2}
 \\ &\qquad  \left(\mathrm{e}^{\frac{i}{\hbar} E_\mathrm{B}(q)t} n(E_\mathrm{B}(q)) + \mathrm{e}^{-\frac{i}{\hbar} E_\mathrm{B}(q)t} [n( E_\mathrm{B}(q))+1] \right)
 \end{split}
 \end{align}
 where we use the Bose-Einstein occupation function of the bath 
 \begin{align}
 	n(E) = \frac{1}{\mathrm{e}^{E/k_B T}-1}.
 \end{align}

We take the continuum limit for the bath sum over~$\vec{q}$, 
$\frac{(2\pi)^3}{V} \sum_{\vec{q}} \rightarrow \int\mathrm{d}^3 q$,
and define the momentum-resolved spectral density
 \begin{align}
 	\tilde{J}(E, {q}) = (u_q-v_q)^2   {n_\mathrm{B}}  \left[ \delta(E-E_\mathrm{B}(q))- \delta(E+E_\mathrm{B}(q))\right],
	\label{eq:Jtilde}
 \end{align}
 which is anti-symmetrized, $\tilde{J}(-E,q)=-\tilde{J}(E,q)$, to allow for a more compact notation in the following.
 With this we may write
 \begin{align}
 \begin{split}
&\langle \tilde{B}(\vec{r},t) \hat B(\vec{r}^{\, \prime}) \rangle_\mathrm{B} \\
& = \frac{1}{(2\pi)^3} \int\! \mathrm{d}^3 q {\mathrm{e}^{-i\vec{q} (\vec{r}- \vec{r}^\prime)}} \! \int_{-\infty}^{\infty} \mathrm{d}E \mathrm{e}^{\frac{i}{\hbar} Et}  \tilde{J}(E, {q})n(E).
 \end{split}
\end{align}
Now we exchange the integrals over $r$ and $q$, identify 
the momentum-space representation of the coupling matrix elements $v^{(K)}_{\alpha \beta}(\vec{q}) ={(2\pi)}^{-3/2}   \int_{r} \mathrm{e}^{-i\vec{q} \vec{r}} v^{(K)}_{\alpha \beta}(\vec{r})$, with coupling matrix elements $v^{(K)}_{\alpha \beta}(\vec{r})$ [cf.~Eq.~\eqref{eq:v-floq}] 
and define the factor 
 \begin{align}
 {f}^{(K)}_{\alpha \beta}(q)=  \int_{\vert\vec{q}^{\, \prime}\vert=q} \!\mathrm{d}^3q'  \vert v^{(K)}_{\alpha \beta}(\vec{q}^{\, \prime})\vert^2
 \end{align}
which integrates the coupling matrix over all vectorial momenta with magnitude $q$. 

This allows us to express the rates as
\begin{align}
	R_{\alpha \beta} = \frac{2\pi \gamma^2}{\hbar} \sum_{K\in \mathbb{Z}}\int_{0}^\infty \!\mathrm{d}q {f}^{(K)}_{\alpha \beta}(q)  \tilde{J}(\Delta^{(K)}_{\alpha \beta}, {q})n(\Delta^{(K)}_{\alpha \beta}).
	\label{eq:rates-general-2}
\end{align}
Performing the integral over $q$ finally yields
\begin{align}
	R_{\alpha \beta} =\frac{2\pi \gamma^2}{\hbar} \sum_{K\in \mathbb{Z}} {f}^{(K)}_{\alpha \beta}\!\!\left(q_\mathrm{B}(\Delta^{(K)}_{\alpha \beta})\right) {J}(\Delta^{(K)}_{\alpha \beta} ) n(\Delta^{(K)}_{\alpha \beta} ),
\end{align}
with 
spectral density
\begin{align}
{J}(E) = \mathrm{sgn}(E) \frac{n_\mathrm{B}}{2}  \frac{q_\mathrm{B}(E)}{\sqrt{E^2 + G^2}}.
	\label{eq:J-K}
 \end{align}
 Here enters the momentum 
 \begin{align}
 q_\mathrm{B}(E)= \frac{\sqrt{2m_\mathrm{B}}}{\hbar} \left(\sqrt{E^2+G^2}-G\right)^{1/2}
  \end{align}
  of a Bogoliubov phonon of energy $E$. 
  Note that for small energies $E \ll G$ we find ohmic behavior $J(E)\propto E$, 
  while for $E \gg G$ the spectral density decays again,
 $J(E)\propto E^{-1/2}$.

In the special case of the immersed one-dimensional optical lattice that we consider, we introduce the Wannier basis $\lbrace \ket{i}\rbrace$ in Eq.~\eqref{eq:v-floq}, $u_{\alpha}(\vec{r},t) = \sum_{i=1}^{M} \braket{i}{u_{\alpha}(t)} w_i(\vec{r})$, with Wannier wavefunction  $w_i(\vec{r})= \braket{\vec{r}}{i}$ at site $i$,  to find the coupling matrix element
 \begin{align}
 	v^{(K)}_{\alpha \beta}(\vec{q}) \approx \int_0^{\mathcal{T}}\! \mathrm{d}t  \intd{^3r} \frac{\mathrm{e}^{-i\vec{q} \vec{r}-iK\omega t}}{(2\pi)^{3/2}\mathcal{T}}  \sum_{i} u_{\alpha,i}(t)^*u_{\beta,i}(t)\vert w_i(\vec{r})\vert^2. 
\end{align}
with $u_{\alpha,i}(t)= \braket{i}{u_{\alpha}(t)}$ and we neglect all contributions from off-site Wannier orbitals $w_i(\vec{r})^*w_j(\vec{r}) \approx \delta_{ij} \vert w_i(\vec{r})\vert^2$.
Performing both integrals, we find
 \begin{align}
 	{f}^{(K)}_{\alpha \beta}(q) =  \sum_{rs} \sum_{ij} u^{(r)*}_{\alpha,i} u^{(r+K)}_{\beta,i}  u^{(s)}_{\alpha,j} u^{(s+K)*}_{\beta,j}  I_{ij}(q),
 \end{align}
 with the $r$-th Fourier component $u^{(r)}_{\alpha,i}= \braket{i}{u_{\alpha}}^{(r)}$ of Floquet mode $\alpha$, that we obtain from solving Eq.~\eqLattfloq in the main text and  
 \begin{align}
 &I_{ij}(q)\\&=\frac{1}{(2\pi)^3}  \int_{\vert\vec{q}^{\, \prime}\!\vert=q} \!\!\mathrm{d}^3q^{\, \prime}\!  \intd{^3r}\intd{^3r'} \mathrm{e}^{-i\vec{q}^{\, \prime} \!(\vec{r}- \vec{r}^{\, \prime})} \vert w_i(\vec{r})\vert^2 \vert w_j(\vec{r}^{\, \prime})\vert^2\\
&=\frac{q \mathrm{e}^{-\frac{1}{2} q^2 d_T^2}}{(2\pi)^2} \int_{-q}^q\mathrm{d}p \,  \mathrm{e}^{-i{p}(x_i-x_j)}  \mathrm{e}^{-\frac{1}{2} p^2 (d_L^2-d_T^2)}.
\label{eq:I-integral-2}
  \end{align}
This was derived by approximating the Wannier functions with harmonic oscillator ground states 
 $w_i(x) \approx \varphi^\mathrm{HO}_0(x-x_i)$ with $x_i = i\cdot a$ effective frequency 
$\Omega_L=2\sqrt{V_0 E_R}/\hbar$. 
Also $d_i=\sqrt{\hbar/m\Omega_i}$ denotes the harmonic oscillator length in the lattice ($i =L$) and the transverse ($i = T$) direction. 
Since the integral over $p$ is hard to evaluate in general, we restrict us for practical reasons to the case where $d_L=d_T$, so that we find 
 \begin{align}
I_{ij}(q)=\left(\frac{q}{2\pi}\right)^2 \mathrm{e}^{-\frac{1}{2} q^2 d_T^2} 2 \mathrm{sinc}\!\left[q(x_i-x_j)\right].
  \end{align}
 Note that we expect similar results for the dynamics also in the general case where $d_L \approx d_T$,  as long as all $q$ fulfill $q\ll \vert d_L^2-d_T^2\vert^{-1/2}$. This can be seen from approximating $\mathrm{e}^{-\frac{1}{2} p^2 (d_L^2-d_T^2)}\approx1$ in the integral of  Eq.~\eqref{eq:I-integral-2}, which is a good approximation for such values $q$.

 In the case where $d_L\gg d_T$, generally we will find $q\gg \vert d_L^2-d_T^2\vert^{-1/2}$ (apart from $q=0$), so a good approximation in this case is (not used in this paper)
 \begin{align}
 I_{ij}(q)&\approx \frac{q \mathrm{e}^{-\frac{1}{2} q^2 d_T^2}}{(2\pi)^2} \int_{-\infty}^\infty\mathrm{d}p \,  \mathrm{e}^{-i{p}(x_i-x_j)}  \mathrm{e}^{-\frac{1}{2} p^2 (d_L^2-d_T^2)}\\
 &=\frac{q \mathrm{e}^{-\frac{1}{2} q^2 d_T^2}}{(2\pi)^{3/2}}  \mathrm{e}^{-\frac{1}{2} \frac{(x_i-x_j)^2}{d_L^2-d_T^2}},
\label{eq:I-integral-app-2}
  \end{align}
so $I_{ij}(q)$ becomes tightly localized in space.}